\documentstyle[12pt]{article}
\def\journal#1, #2, #3, #4 { {\sl #1~}{\bf #2~}(#3) #4 }

\def\mpl{\journal Mod. Phys. Lett., }

\def\prl{\journal Phys. Rev. Lett., }

\def\cmp{\journal Comm. Math. Phys., }

\def\np{\journal Nucl. Phys., }

\def\pl{\journal Phys. Lett., }

\catcode`\@=11
\def\marginnote#1{}
\newcount\hour
\newcount\minute
\newtoks\amorpm
\hour=\time\divide\hour by60
\minute=\time{\multiply\hour by60 \global\advance\minute
by-\hour}\edef\standardtime{{\ifnum\hour<12
\global\amorpm={am}%
        \else\global\amorpm={pm}\advance\hour by-12 \fi
        \ifnum\hour=0 \hour=12 \fi
        \number\hour:\ifnum\minute<10
0\fi\number\minute\the\amorpm}}
\edef\militarytime{\number\hour:\ifnum\minute<10
0\fi\number\minute}

\def\draftlabel#1{{\@bsphack\if@filesw {\let\thepage\relax
   \xdef\@gtempa{\write\@auxout{\string
      \newlabel{#1}{{\@currentlabel}{\thepage}}}}}\@gtempa
   \if@nobreak \ifvmode\nobreak\fi\fi\fi\@esphack}
        \gdef\@eqnlabel{#1}}
\def\@eqnlabel{}
\def\@vacuum{}
\def\draftmarginnote#1{\marginpar{\raggedright\scriptsize\tt#1}}
\def\draft{\oddsidemargin -.5truein
        \def\@oddfoot{\sl preliminary draft \hfil
        \rm\thepage\hfil\sl\today\quad\militarytime}
        \let\@evenfoot\@oddfoot \overfullrule 3pt
        \let\label=\draftlabel
        \let\marginnote=\draftmarginnote

\def\@eqnnum{(\theequation)\rlap{\kern\marginparsep\tt\@eqnlabel}%
\global\let\@eqnlabel\@vacuum}  }

\def\numberbysection{\@addtoreset{equation}{section}
        \def\theequation{\thesection.\arabic{equation}}}

\def\underline#1{\relax\ifmmode\@@underline#1\else
 $\@@underline{\hbox{#1}}$\relax\fi}

\catcode`@=12
\relax
\numberbysection
\def\fin{\end{document}}
\def\beq{\begin{equation}}
\def\eeq{\end{equation}}
\def\beqa{\begin{eqnarray}}
\def\eeqa{\end{eqnarray}}
 \def\nnn{\nonumber \\}
\def\sqr#1#2{{\vcenter{\vbox{\hrule height.#2pt
\hbox{\vrule width.#2pt height#1pt \kern#1pt
\vrule width.#2pt}
\hrule height.#2pt}}}}

\def\pish{{\pi\over h}}

\def\Je{J^e}
\def\Jeb{{\overline J}^e\, \!}

\def\Jne#1 {J_{#1}^e\, \!}
\def\Jneb#1 {{\overline J}_{#1}^e\, \!}
\def\Jnep#1 {J_{#1}'\, \!^e\, \!}
\def\Jnebp#1 {{\overline J}_{#1}'\, \!  \!^e\, \!}

\def\Jehat{{\widehat J}^e}
\def\hhat{{\widehat h}}

\def\Jhat{{\widehat J}}

\def\Nhat{{\widehat N}}

\def\Nhat{{\widehat N}}

\def\rhat{{\widehat r}}

\def\qhat{{\widehat q}}

\def\varpib{{\overline \varpi}}

\def\phat{{\widehat p}}

\def\zb{{\bar z}}

\def\pb{{\bar p}}
\def\pbhat{\widehat{{\bar p}}} 
\def\Vb{{\overline V}}

\def\Jgen#1 {  {\underline J_{#1}} }
\def\Jgenp#1 #2 {(J_{#1}+{#2},\Jhat_{#1})}
\def\Jgenm#1 #2 {(J_{#1}-{#2},\Jhat_{#1})}
\def\Jg#1 {J_{#1},\Jhat_{#1}}
\def\Jgp#1 #2 {J_{#1}+{#2},\Jhat_{#1}}
\def\Mgen#1 {{\underline M_{#1}}}

\def\produit#1,#2,#3,#4 {P\Bigl ( [{#1},{#2}]\otimes\{{#3}\},{#4}\Bigr )}
\def\produitscript#1,#2,#3,#4 {P\Bigl (
[{\scriptstyle{#1},{#2}}]\otimes\{{\scriptstyle{#3}}\},{#4}\Bigr )}
\def\pprod#1,#2,#3,#4,#5 {P\Bigl ( [{#1},{#2}]\otimes[{#3},{#4}],{#5}\Bigr )}
\def\pprodscript#1,#2,#3,#4,#5 {P\Bigl (
[{\scriptstyle{#1},{#2}}]\otimes[{\scriptstyle{#3},{#4}}],{#5}\Bigr )}

\def\fusV#1,#2,#3,#4,#5,#6 {f_V(
\Jgen{#1} ,
\Jgen{#2} ,
\Jgen{#3} ,
\Jgen{#4} ,
\Jgen{#5} ,
\Jgen{#6} )}

\def\brdV#1,#2,#3,#4,#5,#6 {b_V(
\Jgen{#1} ,
\Jgen{#2} ,
\Jgen{#3} ,
\Jgen{#4} ,
\Jgen{#5} ,
\Jgen{#6} )}

\def\fusxi#1,#2,#3 {f_\xi (\Jgen{#1} ,
\Mgen{#1} ,
\Jgen{#2} ,
\Mgen{#2} ,
\Jgen{#3} )}

\def\gaghat{{\hat {\bigl \{}}}
\def\gadhat{{\hat {\bigr \}}}}

\def\bverthat{{\hat {\bigl \vert}}}

\def\sixjxi#1,#2,#3,#4,#5,#6 {{\left\{\left . \!\! \,^{#1}_{#2}
\,^{#3}_{#4} \right | \!\, ^{#5}_{#6}\right\}}}
\def\sixje#1,#2,#3,#4,#5,#6 {{\left\{\left\{\left . \!\! \, ^{#1}_{#2}
\, ^{#3}_{#4} \right | \!\, ^{#5}_{#6}\right\}\right\}}}
\def\sixjxihat#1,#2,#3,#4,#5,#6 {{{\gaghat\left . \!\! \, ^{#1}_{#2}
\, ^{#3}_{#4} \right | \!\, ^{#5}_{#6}\gadhat}}}
\def\sixjehat#1,#2,#3,#4,#5,#6 {{\gaghat\gaghat\left . \!\! \, ^{#1}_{#2}
\, ^{#3}_{#4} \right | \!\, ^{#5}_{#6}\gadhat\gadhat}}

\def\Jehat {{\widehat J^e}}

\def\epsilonhat{{\widehat \epsilon}}
\def\pb{\overline p}

\begin{document}
\begin{titlepage}

\begin{flushright}

LPTENS--96/01, \\
hep-th/9601034, \\
January 1996
\end{flushright}

\vglue 2.5  true cm
\begin{center}
{\large \bf
ON THE LIOUVILLE COUPLING CONSTANTS\\}   
\vglue 1.5 true cm
{\bf Jean-Loup~GERVAIS}\\
\medskip
\medskip
{\footnotesize Laboratoire de Physique Th\'eorique de
l'\'Ecole Normale Sup\'erieure\footnote{Unit\'e Propre du
Centre National de la Recherche Scientifique,
associ\'ee \`a l'\'Ecole Normale Sup\'erieure et \`a
l'Universit\'e
de Paris-Sud.},\\
24 rue Lhomond, 75231 Paris CEDEX 05, ~France}.
\end{center}
\vfill
\begin{abstract}
\baselineskip .4 true cm
\noindent
{\footnotesize
For  the general operator
product algebra coefficients derived by Cremmer Roussel Schnittger  and
the present author with  (non negative) 
integer screening numbers, the coupling constants
determine the factors additional to the quantum group 6j symbols.
 They are given by  path independent products over 
a two dimensional lattice in the zero mode space. 
It  is shown that the ansatz for the three point function
of Dorn-Otto and  Zamolodchikov-Zamolodchikov  precisely defines the
corresponding flat lattice connection, so that it
does give a natural generalization  of these  coupling constants  to
continuous screening numbers. The consistency of the  restriction to 
integer 
screening charges is reviewed, and shown to be linked with the
orthogonality of the (generalized)  6j symbols.  Thus  extending this 
last relation is the key to general screening numbers.  
 }
\end{abstract}
\vfill
\end{titlepage}
\section{Introduction}  
One outcome of refs.\cite{CGR1,CGR2,GS1,GS2,GR1} was the general
expression for the fusion and braiding matrices  of 
the general chiral operators noted $V^{(\Je )}(z)$, which are chiral components of 
the quantum Liouville exponentials. We  follow the notational  conventions of these 
works. Calling $h$ the quantum group deformation parameter, the central charge is 
$C_L=1+6({h\over \pi}+{\pi\over h}+2)$, 
The notation $\Je$ 
characterizes a primary field with associated (rescaled) Liouville 
momentum $\varpi_{\Je}=\varpi_0+2\Je$, where $\varpi_0=1+\pi/h$ is one of the 
$sl(2)$ invariant vacua. Its weight is $\Delta(\Je)=(\varpi_0^2-\varpi_{\Je}^2)h/4\pi$. 
The discussion was carried out for arbitrary continuous $\Je$, but for the degenerate 
fields, one has $\Je=J+\Jhat \pi/h$, such that $2J+1$ and $2\Jhat+1$ are positive integers 
which caracterize the finite dimensional quantum group representations. Thus $\Je$ is thought of 
as the effective spin.    
We denote by  ${\cal P}_{\Je }$, the projector over the corresponding Verma 
module.  
The full fusion-equation reads\cite{GR1}  
$$
{\cal P}_{\Je_{123} }
V^{(\Je_1)}(z_1) {\cal P}_{\Je_{23}} V^{(\Je_2)}(z_2)
{\cal P}_{\Je_3 }
=\sum _{\Je_{12}}
{g_{\Je_1\Je_2}^{\Je_{12}}\
g_{\Je_{12}\Je_3}^{\Je_{123}}
\over
g_{\Je_2\Je_3}^{\Je_{23}}\
g_{{\Je_1}\Je_{23}}^{\Je_{123}}
}
\left\{\left\{ ^{\Je_1}_{\Je_3}\,^{\Je_2}_{\Je_{123}}
\right. \left |^{\Je_{12}}_{\Je_{23}}\right\}\right\}_q
\left\{\left\{  ^{\Jehat_1}_{\Jehat_3}\,^{\Jehat_2}_{\Jehat_{123}}
\bigr. \bverthat\, ^{\Jehat_{12}}_{\Jehat_{23}}\right\}\right\}_{\qhat}
$$
\beq
\sum _{\{\nu_{12}\}}
{\cal P}_{\Je_{123} }
V ^{(\Je_{12},\{\nu_{12}\})}(z_2)
{\cal P}_{\Je_3 }<\!\varpi_{\Je_{12}},{\{\nu_{12}\}} \vert
V ^{(\Je_1)}(z_1-z_2) \vert \varpi_{\Je_2} \! >.
\label{fusion}
\eeq 
Note that the last term is a c number (a matrix element) which is a book keeping device 
to handle all descendents at once. They are characterized, abstractly by a multi-index 
noted $\{\nu\}$.   
The $V$ fields are normalized such that
\beqa 
<\!\varpi_{L^e} \vert
V ^{(\Je)}(0) \vert \varpi_{K^e} \! > =1&\hbox{if} & 
\Je+K^e-L^e=p+\phat\pi/h,\quad   
p\in {\cal Z}, \> \phat\in {\cal Z},
\label{norm}\\
<\!\varpi_{L^e} \vert
V ^{(\Je)}(0) \vert \varpi_{K^e} \! > =0&&\hbox{otherwize}  
\label{screen}
\eeqa
where ${\cal Z}$ is the set of non negative integers. 
The symbol $g_{J^e K^e}^{L^e}$ stands for the coupling constants which are 
the central point of the present note. They involve the contributions which are not 
solely determined by the quantum group symmetry, in contrast with the 6j symbols. 
The sum over $\{\nu_{12}\}$ represents the 
summation over arbitary states\footnote{to simplify the formulae we
assume that they are orthonormalized.}  
 of the Verma module with momentum $\varpi_{\Je_{12}}$. 
This equation was derived for the most general case where the $\Je$'s are 
arbitrary continuous variables with the restriction that  condition 
Eq.\ref{screen} 
be obeyed by each of the four $V$ operators which appear in the fusion equation 
\ref{fusion}, so that 
\beqa
&\Je_1+\Je_{23}-\Je_{123}=p_{1,23}+\pish \phat_{1,23}, &p_{1,23},\, \phat_{1,23} \in {\cal Z}
\nnn
&\Je_2+\Je_3-\Je_{23}=p_{2,3}+\pish \phat_{2,3}, &p_{2,3},\, \phat_{2,3} \in {\cal Z}
\nnn
&\Je_{12}+\Je_3-\Je_{123}=p_{12,3}+\pish \phat_{12,3},&p_{12,3},\, \phat_{12,3} \in {\cal Z}
\nnn
&\Je_1+\Je_2-\Je_{12}=p_{1,2}+\pish \phat_{1,2}, &p_{1,2},\, \phat_{1,2} \in {\cal Z}. 
\label{1PI}
\eeqa
From the viewpoint of Coulomb gas, the $p$'s are the screening numbers. 
 Thus, there is a consistent operator product algebra where 
all these numbers are non negative  integers.     
  The symbols between double braces are the corresponding 
generalized 6j symbols associated with the two quantum group parameters 
$h$, $\hhat$ ($q=e^{ih}$, $\qhat =e^{i\hhat}$), themselves 
related to the two screening charges $\alpha_\pm$ by $h=\pi\alpha_-^2/\pi$, 
$\hhat=\pi\alpha_+^2/\pi$. In the 6j symbols 
 associated with $\qhat$, we have introduced 
the convenient notation $\Jehat =\Je h/\pi$ which makes the symmetry between 
$h$ and $\hhat$ more explicit.   
The sum over $\Je_{12}$ runs over all the values of $\Je_{12}$ allowed
by the four screening number conditions.
It can be viewed as a double sum on $p_{1,2}$, $\phat_{1,2}$, such that 
$p_{1,2}+\phat_{1,2}\pi/h=\Je_1+\Je_2-\Je_{12}$. Thus it is a summation 
over 
non negative 
 integers. By construction, the fusion  formula  only involves the coupling 
constants $g_{J^e K^e}^{L^e}$, with the restriction that 
$J^e+K^e-L^e=p+\phat \pi/h$, where $p$, and $\phat$ are non negative
 integers. 

We recall the braiding equation as well.
It was derived in ref.\cite{GS1} for one half of the algebra
and in ref.\cite{GS2}
for the full algebra. 
It can be deduced from the fusion
by the three leg symmetry of the vertices\cite{MS,CGR1,GR1}:
\beq
<\varpi_{12} |  V^{(\Je_1)}
|\varpi_2>
= e^{i\pi(\Delta(\Je_1)+\Delta(\Je_2)-\Delta(\Je_{12}))}
<\varpi_{\Je_{12}} |  V^{(\Je_2)} |\varpi_1>. 
\label{sym}
\eeq
This  gives
$$
{\cal P}_{\Je_{123} }
V^{(\Je_1)}(z_1) {\cal P}_{\Je_{23}}
V^{(\Je_2)}(z_2) 
{\cal P}_{\Je_3 }
=
\sum_{\Je_{13}} e^{\pm i\pi (\Delta(\Je_{123})+\Delta(\Je_3)
-\Delta(\Je_{23})-\Delta(\Je_{13}))} \times
$$
\beq
{g_{\Je_1 \Je_3}^{\Je_{13}} g_{\Je_{13} \Je_2}^{\Je_{123}} \over
g_{\Je_2 \Je_3}^{\Je_{23}} g_{\Je_1 \Je_{23}}^{\Je_{123}}}
\left\{\left\{
^{\Je_1}_{\Je_2}\,^{\Je_3}_{\Je_{123}}
\right. \left |^{\Je_{13}}_{\Je_{23}}\right\}\right\}_q
\left\{\left\{
\,^{\Jehat_1}_{\Jehat_2}\,^{\Jehat_3}_{\Jehat_{123}}
\bigr. \bverthat\, ^{\Jehat_{13}}_{\Jehat_{23}}\right\}\right\}_{\qhat} 
{\cal P}_{\Je_{123} }
V ^{(\Je_2)}(z_2) {\cal P}_{\Je_{13}}
V^{(\Je_1)}(z_1)
{\cal P}_{\Je_3 }
\label{braiding}
\eeq
where again the sum over $\Je_{13}$ is to be understood as a double sum.

\section{The  coupling constant from a flat connection}  
The general expression of the coupling constants was given in refs.\cite{CGR1,GR1} 
under the form 
\beq
g_{\Je_1,\Je_2}^{\Je_{3}}= g_0^p \hat g_0^{\phat}  
{H_{p\phat}(\varpi_{\Je_1})H_{p\phat}(\varpi_{\Je_2})
H_{p\phat}(-\varpi_{\Je_{3}})
\over{H_{p\phat}(\varpi_{p/2,\phat /2})}}
\label{ggen}
\eeq
with $\varpi_{p/2,\phat /2}=\varpi_0+p+\phat \pi/h$, $\Je_1+\Je_2-\Je_{12} =
p+\phat \pi/h$, $p,\, \phat \in {\cal Z}$,  and 
\beq
H_{p\phat}(\varpi) =
{
\prod_{r=1}^p \sqrt{F((\varpi -r) h/\pi )}
\prod_{\rhat =1}^{\phat}
\sqrt{F(\varpi - \rhat \pi /h )}
\over
\prod_{r=1}^p \prod_{\rhat =1 }^{\phat }
\left ( \varpi \sqrt{h/\pi} 
-r \sqrt{h/\pi} -\rhat \sqrt{\pi /h}\right ) }
\label{H}.
\eeq
The constants $g_0$ and $\hat g_0$ are arbitrary. The function $F$ is defined by 
\beq
F(z)={\Gamma(z)\over \Gamma(1-z)}. 
\label{fdef}
\eeq
 By absorbing the denominator factors, 
 an equivalent form was given:   
$$
H_{p\phat}(\varpi) =
\prod_{i=1}^{n-1}
\left\{
F\left[\varpi-N_i-(\Nhat_i+{1+\epsilonhat_i\over2}){\pi\over h}\right]
\left({-\pi\over h}\right)^{N_i}
\right\}^{\epsilonhat_i/2}
$$
\beq
\left\{
F\left[{h\over \pi}\varpi-\Nhat_i-(N_i+{1+\epsilon_i\over2}){h\over \pi}\right]
\left({-h\over\pi}\right)^{\Nhat_i}
\right\}^{\epsilon_i/2}
\label{H2}
\eeq
where the ${(N_i,\Nhat_i),i=1...n}$ describe an arbitrary planar path
going from $(0,0)=(N_1,\Nhat_1)$ to $(p,\phat)=(N_n,\Nhat_n)$.
The allowed elementary steps $(\epsilon_i,\epsilonhat_i)\equiv
(N_{i+1}-N_i,\Nhat_{i+1}-\Nhat_i)$
are $(0,\pm 1)$ and $(\pm 1,0)$. 
Of course, this expression only makes sens 
if $p$ and $\phat$ are integers. However, it is not restricted to the case 
where they are  positive (more  on this below). One sees that it is given by a 
sort of Wilson product  of a discrete flat connection in a two dimensional 
square lattice with spacings $1$ and $\pi/h$. The point of the present note is 
to show that the function $\Upsilon$ introduced in refs.\cite{DO,ZZ} 
precisely allows us  to 
integrate explicitly the above Wilson integral. This will  give a natural  interpolation for 
the  connection 
away from the discrete lattice, which will 
define $g$ for arbitrary continuous $p$ and $\phat$. 

The derivation goes as follows. Recall the definition of refs.\cite{DO,ZZ}:
\beq
F(bx)={\Upsilon(x+b)\over \Upsilon(x)} b^{2bx-1},\quad 
F({1\over b}x)={\Upsilon(x+{1\over b})\over \Upsilon(x)} \left({1\over b}\right)^{2x/b-1}
\label{upsilondef}
\eeq
To clarify the gauge analogy, consider an abelian  flat connection in two dimensions 
$\vec A =\vec \nabla \Lambda$. We put it on a square lattice with spacings 
$b$ and $1/b$ by introducing the Wilson link variables 
\beq
U_{\vec r, \vec \mu}=e^{\int _{\vec r}^ {\vec r+\vec \mu} \vec A  d\vec x}=
e^{\Lambda(\vec r+\vec \mu)-\Lambda(\vec r)},  
\label{Udef}
\eeq
where $\vec r$ is a point on the lattice (with coordinates 
$(nb,\, m/b)$ $n$, $m$ integers), 
and $\vec \mu$ has coordinates $(b,\, 0)$ or $(0,\, 1/b)$.  
Of course, by construction, the Wilson line product 
along any path 
on the lattice  
only depends upon the initial and final points:
\beq
\prod_{i=0}^N  U_{\vec r_i, \vec \mu_i}=e^{\Lambda(\vec r_N+\vec \mu_N)-\Lambda(\vec r_0)}
\label{proddef}
\eeq 
 To make contact with 
the previous quantum group formulae, we let 
\beq
b=\sqrt{h\over \pi}\quad, \varpi={1\over b} x, 
\label{bdef}
\eeq 
and rewrite Eq.\ref{H2} as   
$$
H_{p\phat}(\varpi) =
\prod_{i=1}^{n-1}
\left\{
F\left[{1\over b}\left ( x-bN_i-{1\over b}(\Nhat_i+{1+\epsilonhat_i\over2})\right)\right]
\left({-1\over b^2}\right)^{N_i}
\right\}^{\epsilonhat_i/2}
$$
\beq
\left\{
F\left[b\left(x-{1\over b}\Nhat_i-(N_i+{1+\epsilon_i\over2})b\right)\right]
\left({-b^2}\right)^{\Nhat_i}
\right\}^{\epsilon_i/2}. 
\label{H3}
\eeq
Clearly, the values of $x$ which appear are of the form $x=x_0+nb+m/b$, $n$ $m$ integers, $x_0$ 
fixed continuous constant. 
If $b$ is irrational, 
this uniquely defines a point in the two dimensional square lattice introduced above. Thus we 
have a discrete flat connection of the general type mentioned above. The basic point is that 
Eqs.\ref{upsilondef} simply define the corresponding gauge function $\Lambda$. This is easily 
seen by using Eqs.\ref{upsilondef} to integrate the product \ref{H3}. Since the result does not 
depend upon the path, one may take the simplest rectangular one. Then one finds
\beq
H_{p,\, \phat}(\varpi)={\sqrt{\Upsilon(x)}\over \sqrt{\Upsilon(x-pb-\phat {1\over b})}}
b^{(2bx-1-b^2(p+1))p/2} \left({1\over b}\right)^{(2x/b-1-(\phat+1)/b^2)\phat/2}
\label{H4}
\eeq
One sees that up to the last two terms we have obtained an equation of the type Eq.\ref{proddef}, 
with $\Lambda=\ln (\Upsilon (x))/2$. Of course, once it is derived it give a natural 
generalisation of the flat connection to the continuum -- reversing the 
discretization performed by writing Eq.\ref{Udef}.  The next question concerns possible 
zeros or poles of $\Upsilon$. Their positions are  easily seen by looking at possible 
singularities in the path formulation Eq.\ref{H3}. It follows from Eqs.\ref{fdef}, 
\ref{upsilondef} that 
\beq
{\Upsilon(x+b)\over \Upsilon(x)}
\left\{ \begin{array}{l}
=\infty\> \hbox{if} \> x=-n/b,\quad n\in {\cal Z};  \nnn
=0 \> \hbox{if} \> x=(n+1)/b,\quad n\in {\cal Z}; \nnn
\hbox{ is regular otherwize}
\end{array}
\right.
\eeq
 \beq
{\Upsilon(x+{1\over b})\over \Upsilon(x)}
\left\{ \begin{array}{l}
=\infty\> \hbox{if} \> x=-nb\quad n\in {\cal Z}; \nnn
=0 \> \hbox{if} \> x=(n+1)b,\quad n\in {\cal Z}; \nnn
\hbox{ is regular otherwize;}
\end{array}
\right.
\eeq
This agrees with the fact that $\Upsilon(x)$ vanishes iff $x=-nb-m/b$, 
and $x=(n+1)b+(m+1)/b$ with 
 $n$ and $m$ both non-negative integers.  

  Returning to Eq.\ref{ggen}, we next deduce the expression 
for the coupling constant, namely,
$$
g_{\Je_1,\Je_2}^{\Je_{3}}= 
\left(g_0b^{-(1+b^2)}\right)^p \left ((\hat g_0/ b)^{-(1+1/b^2)}\right)^{\phat}\times
$$
\beq 
{\sqrt{\Upsilon(x_1)}\sqrt{\Upsilon(x_2)}\sqrt{\Upsilon(-x_3)}\sqrt{\Upsilon(b+{1\over b})}\over 
\sqrt{\Upsilon(x_1+pb+\phat{1\over b})}\sqrt{\Upsilon(x_2+pb+\phat{1\over b})}
\sqrt{\Upsilon(-x_3+pb+\phat{1\over b})} 
\sqrt{\Upsilon((p+1)b+(\phat+1){1\over b})}},
\label{coupl}  
\eeq
where we have let $x_i=b\varpi_{\Je_i}$, so that
\beq
x_1+x_2-x_3=(2p+1)b+(2\phat+1)/b. 
\label{screenings}
\eeq 
Several comments are in order. Let us begin by  discussing  the properties of the expression 
just derived as it is (we shall connect with the earlier expressions of 
refs.\cite{DO}, 
\cite{ZZ} last).  First, in the most restricted case, namely when $\Je_i=J_i+
\Jhat_i \pi/h$, $2J_i\in {\cal Z}$, $2\Jhat_i\in {\cal Z}$. All the terms is the 
numerator and in denominator vanish, so that this formula is not very useful. 
The expression is perfectly finite however, and is better derived from the previous 
product relation \ref{H} or \ref{H2}. Second, consider the case where 
 the spins are continuous, but $p$ and $\phat$ 
are still  integers. If $p$ and $\phat$ are both positive or both non-positive, the 
vanishing of $\Upsilon((p+1)b+(\phat+1){1\over b})$ in the denominator compensates the 
vanishing of  $\Upsilon(b+{1\over b})$ in the numerator and 
 the ratio is still finite. Third, on the contrary,  in the mixed situation,  
Eq.\ref{coupl} gives zero since it contains  $\sqrt{\Upsilon(b+1/b)}$ 
as the only vanishing term in the numerator. Fourth, if everything is continuous, 
Eq.\ref{coupl} gives zero, for the same reason. 

In the earlier discussions\cite{DO}\cite{ZZ}, one modifies the 
above expression, and replaces  the vanishing term $\Upsilon(b+1/b)$ by its derivative which 
is non zero. As a result, the new expression blows up when Eq.\ref{coupl} is finite, and is finite 
when Eq.\ref{coupl} gives a vanishing result. This explains why the cases discussed in 
refs.\cite{CGR1}--\cite{GR1} appear as poles in 
the work of \cite{DO},\cite{ZZ}.   A similar situation is 
encountered in  the heuristic expressions  proposed  in ref.\cite{GL} 
 of amplitudes with continuous screnning which are   proportional to 
gamma functions\footnote{
This is really the case only if the two screening
charges are dealt with independently contrary to ref.\cite{GL}, as was
done for instance in ref.\cite{D}.} 
 whose arguements  are the opposite of the screening
numbers, which one may factor out at will. 

\section{The Liouville n-point functions.}
Since this question has attracted much attention lately, it is worthwile summarizing 
how they arise in the present context. We shall deal with the   
conformal boostrap solution based on the fusing and braiding relations recalled above 
where the screening numbers are consistently kept integers.  
On the sphere, the  n-point functions of chiral operators may be calculated from 
the matrix element
\beq
G^{(n,\, V)}_{\varpi_1, \varpi_2,\, \cdots ,\, \varpi_n}(z_1,\, z_2, \cdots ,\, z_n)
=<-\varpi_0| V^{(\Je_n)}(z_n) V^{(\Je_{n-1})}(z_{n-1})\cdots V^{(\Je_1)}(z_1)|\varpi_0>, 
\label{nptdef}
\eeq
where 
\beq
\varpi_i=\varpi_0+2\Je_i.
\label{nptdef2}
\eeq
Inserting complete sums over intermediate states, one sees that we may rewrite
$$
G^{(n,\, V)}_{\varpi_1, \varpi_2,\, \cdots ,\, \varpi_n}(z_1,\, z_2, \cdots ,\, z_n)
= \sum_{K^e_1,\, \cdots K^e_{n-1}}
$$
\beq<-\varpi_0| V^{(\Je_n)}(z_n) 
{\cal P}_{K^e_{n-1}}V^{(\Je_{n-1})}(z_{n-1}){\cal P}_{K^e_{n-2}}
\cdots {\cal P}_{K^e_1}V^{(\Je_1)}(z_1)|\varpi_0>, 
\label{nptdef3}
\eeq
As usual, M\"obius invariance is such that it is enough to compute in the limit $z_n\to 0$, 
$z_1\to\infty$. One gets\footnote{ up to a divergent factor in $z_n$ which we drop} 
$$
G^{(n,\, V)}_{\varpi_1, \varpi_2,\, \cdots ,\, \varpi_n}(0,\, z_2, \cdots ,\ z_{n-1},\, \infty)
= \sum_{K^e_2,\, \cdots K^e_{n-2}}
$$
\beq<-\varpi_n| 
V^{(\Je_{n-1})}(z_{n-1}){\cal P}_{K^e_{n-2}}
\cdots {\cal P}_{K^e_2}V^{(\Je_2)}(z_2)|\varpi_1>, 
\label{nptdef4}
\eeq
This fact was actually derived and shown to be consistent with the operator algebra for 
half integer spins only. There may be subtleties in the general case which we shall leave out. 
In any case, the restriction to non negative 
integer values of the screening operators is such that the 
summations over the  intermediate spins is discrete. This is easily seen by recursion since 
$K^e_j$ must be of the form $K^e_j=K^e_{j-1}+\Je_j+p_j+\phat_j\pi/h$ with $p_j$ and $\phat_j$ 
integers. Concerning the  four-point function,  
 Eq.\ref{fusion}, allows us to  re-express it in terms of 
three-point functions as follows
$$
G^{(4,\, V)}_{\varpi_1, \varpi_2,\, \varpi_3,\, -\varpi_4}(0,\, z_2, z_3 ,\, \infty)
=\sum_{\Je_{23}}\left [\sum _{\Je_{12}}\left (
{g_{\Je_3\Je_2}^{\Je_{23}}\
g_{\Je_{1}\Je_{23}}^{\Je_{4}}
\over
g_{\Je_2\Je_1}^{\Je_{12}}\
g_{{\Je_{12}}\Je_{4}}^{\Je_{3}}
}
\left\{\left\{ ^{\Je_3}_{\Je_1}\,^{\Je_2}_{\Je_{4}}
\right. \left |^{\Je_{23}}_{\Je_{12}}\right\}\right\}_q
\left\{\left\{  ^{\Jehat_2}_{\Jehat_1}\,^{\Jehat_3}_{\Jehat_{4}}
\bigr. \bverthat\, ^{\Jehat_{23}}_{\Jehat_{12}}\right\}\right\}_{\qhat}\right.\right)
$$
\beq
\left. \sum _{\{\nu_{23}\}}
G^{(3,\, V)}_{\varpi_1, \varpi_{23},\{\nu_{23}\},\, -\varpi_4}(0,\, z_1, \, \infty)
G^{(3,\, V)}_{\varpi_2, \varpi_{3}, -\varpi_{23}\{\nu_{23}\},}(0,\, z_3-z_1, \, \infty)
\right]
\label{4-3pts}
\eeq
This involves of course  three point functions with descendents 
defined, in general, according to  
\beq
G^{(3,\, V)}_{\varpi_1, \{\nu_{1}\}\varpi_{2},\{\nu_{2}\},\, 
\varpi_3, \{\nu_{3}\}}(0,\, z_2, \, \infty)=
<-\varpi_3, \{\nu_3\}| V^{(\Je_2,\{\nu_2\})}(z_2)|  \varpi_1, \{\nu_1\}>
\label{3ptgen}
\eeq
In this way all n point functions are expressed as discrete sums 
 of three point functions. 
The braiding  equation \ref{braiding} leads to similar relations, albeit with quantum numbers 
put differently:
$$
G^{(4,\, V)}_{\varpi_1, \varpi_2,\, \varpi_3,\, -\varpi_4}(0,\, z_2, z_3 ,\, \infty)
=
$$
$$\sum_{\Je_{13}}\left [\sum _{\Je_{12}}\left (
{g_{\Je_{13}\Je_2}^{\Je_{4}}\
g_{\Je_{1}\Je_{3}}^{\Je_{13}}
\over
g_{\Je_2\Je_1}^{\Je_{12}}\
g_{{\Je_{12}}\Je_{4}}^{\Je_{3}}
}
\left\{\left\{ ^{\Je_3}_{\Je_2}\,^{\Je_1}_{\Je_{4}}
\right. \left |^{\Je_{13}}_{\Je_{12}}\right\}\right\}_q
\left\{\left\{  ^{\Jehat_3}_{\Jehat_2}\,^{\Jehat_1}_{\Jehat_{4}}
\bigr. \bverthat\, ^{\Jehat_{13}}_{\Jehat_{12}}\right\}\right\}_{\qhat} 
 e^{\pm i\pi (\Delta(\Je_{4})+\Delta(\Je_1)
-\Delta(\Je_{12})-\Delta(\Je_{13}))} \right)\right.
$$
\beq
\left .\sum _{\{\nu_{13}\}}
G^{(3,\, V)}_
{\varpi_{13}\{\nu_{13}\},\, \varpi_2\, -\varpi_4}(0,\, z_2, \, \infty)
G^{(3,\, V)}_
{\varpi_{1}, \varpi_3, \, -\varpi_{13} , \{\nu_{13}\}, \varpi_{3},}
(0,\, z_3, \, \infty)
\right]. 
\label{4ptbraid}
\eeq 
In general, there is no way to reexpress the   right hand side in terms of a 
4 point function, since the 
braiding matrix explicitly depends upon $\Je_{13}$. Thus the n point functions of the 
$V$ fields do not have well defined monodromy properties. 
This will not be the case for the expectation values of the Liouville exponentials as we next show. 
According to the  operator expression\cite{GS2} of the Liouville 
exponentials,  they are defined such that 
$$
G^{(n,\, \Phi)}_{\varpi_1, \varpi_2,\, \cdots ,\, -\varpi_n}(0, 0,\, z_2,\zb_2  \cdots ,\,
 \infty ,\infty)=
$$
$$
 \sum_{K^e_2,\, \cdots K^e_{n-2}} \left ( g_{\Je_2 \Je_1}^{K_2^e}g_{\Je_3 K_2^e}^{K_3^e} 
\cdots  g_{\Je_{n-2} K_{n-3}^e}^{K_{n-2}^e}  g_{\Je_{n-1}K_{n-2}^e}^{J_{n}^e}\right)^2  
$$
$$
<\varpi_n| 
V^{(\Je_{n-1})}(z_{n-1}){\cal P}_{K^e_{n-2}}
\cdots {\cal P}_{K^e_2}V^{(\Je_2)}(z_2)|\varpi_1>
$$
\beq 
<\varpi_n| 
\Vb^{(\Je_{n-1})}(\zb_{n-1}){\cal P}_{K^e_{n-2}}
\cdots {\cal P}_{K^e_2}V^{(\Je_2)}(\zb_2)|\varpi_1>, 
\label{nptphi}
\eeq
 Quantities
with a bar denote the other chiral component expressions. 
Let us now return to the braiding of the 4 pt function, which reads 
$$
G^{(4,\, \Phi)}
_{\varpi_1, \varpi_2,\, \varpi_3,\, -\varpi_4}(0, 0,\,
z_2,\zb_2 ,\,
z_2,\, \zb_3,\, \infty ,\infty)=
 \sum_{J^e_{12} } \left ( g_{\Je_2 \Je_1}^{J^e_{12}}
g_{\Je_3
J^e_{12}}^{J_4^e}
\right)^2
$$
\beq
<\varpi_4|
V^{(\Je_{3})}(z_{3}){\cal P}_{J^e_{12}}
V^{(\Je_2)}(z_2)|\varpi_1>
<\varpi_4| 
\Vb^{(\Je_{3})}(\zb_{3}){\cal P}_{J^e_{12}} 
\Vb^{(\Je_2)}(\zb_2)|\varpi_1>. 
\label{4ptphi}
\eeq
 Making use of Eq.\ref{4ptbraid} 
for the two chiralities one gets
$$
G^{(4,\, \Phi)}
_{\varpi_1, \varpi_2,\, \varpi_3,\, -\varpi_4}(0, 0,\, z_2,\zb_2 ,\,
z_3,\, \zb_3,\, \infty ,\infty)=
\sum_{\Je_{13},\overline J^e_{13}}
 g_{\Je_{13}\Je_2}^{\Je_{4}}\ g_{\overline J^e_{13} \Je_2}^{\Je_{4}}
g_{\Je_{1}\Je_{3}}^{\Je_{13}} g_{\Je_{1}\Je_{3}}^{\overline J^e_{13}}\times  
$$
$$
\left [\sum _{\Je_{12}}\left (
\left\{\left\{ ^{\Je_3}_{\Je_2}\,^{\Je_1}_{\Je_{4}}
\right. \left |^{\Je_{13}}_{\Je_{12}}\right\}\right\}_q
\left\{\left\{  ^{\Jehat_3}_{\Jehat_2}\,^{\Jehat_1}_{\Jehat_{4}}
\bigr. \bverthat\, ^{\Jehat_{13}}_{\Jehat_{12}}\right\}\right\}_{\qhat} 
\left\{\left\{ ^{\Je_3}_{\Je_2}\,^{\Je_1}_{\Je_{4}}
\right. \left |^{\overline J^e_{13}}_{\Je_{12}}\right\}\right\}_q
\left\{\left\{  ^{\Jehat_3}_{\Jehat_2}\,^{\Jehat_1}_{\Jehat_{4}}
\bigr. \bverthat\, ^{{\widehat {\overline J}}^e_{13}}_{\Jehat_{12}}\right\}\right\}_{\qhat} 
 \right)\right.  
$$
$$
\sum _{\{\nu_{13}\}}
G^{(3,\, V)}_{\varpi_{13}\{\nu_{13}\},\, \varpi_2,\, -\varpi_4}
(0,\, z_2, \, \infty)
G^{(3,\, V)}_{\varpi_1, \varpi_3,\, -\varpi_{13} , \{\nu_{13}\}} 
(0,\, z_3, \, \infty)
$$
\beq
\left. \sum _{\{\overline \nu_{13}\}}
 G^{(3,\,\Vb)}_{\varpib_{13}\{\overline \nu_{13}\},\, \varpi_2,\, -\varpi_4}
(0,\, \zb_2, \, \infty)
 G^{(3,\, \Vb)}_
{\varpi_1, \varpi_3,\, -\varpib_{13},  \{\overline \nu_{13}\} } 
(0,\, \zb_3, \, \infty)
\right]. 
\label{phibraid1}
\eeq
The basic difference with Eq.\ref{4ptphi} 
is that the summation over the intermediate ($13$) quantum numbers 
is done independently 
for the two chiralities, since it only appears on the right hand side of Eq.\ref{4ptbraid}. 
 The symbol $\varpib_{13}$ is defined as equal to $\varpi_0+2\Jeb_{13}$. However, it was verified 
in ref.\cite{GR1} that, as long as the screening numbers are  
non negative integers, the generalized 6j 
are orthonormal polynomials such that 
\beq
 \sum _{\Je_{12}}\left (
\left\{\left\{ ^{\Je_3}_{\Je_2}\,^{\Je_1}_{\Je_{4}}
\right. \left |^{\Je_{13}}_{\Je_{12}}\right\}\right\}_q
\left\{\left\{  ^{\Jehat_3}_{\Jehat_2}\,^{\Jehat_1}_{\Jehat_{4}}
\bigr. \bverthat\, ^{\Jehat_{13}}_{\Jehat_{12}}\right\}\right\}_{\qhat} 
\left\{\left\{ ^{\Je_3}_{\Je_2}\,^{\Je_1}_{\Je_{4}}
\right. \left |^{\overline J^e_{13}}_{\Je_{12}}\right\}\right\}_q
\left\{\left\{  ^{\Jehat_3}_{\Jehat_2}\,^{\Jehat_1}_{\Jehat_{4}}
\bigr. \bverthat\, ^{{\widehat {\overline J}}^e_{13}}_{\Jehat_{12}}\right\}\right\}_{\qhat} 
 \right)=\delta_{\Je_{13},\overline J^e_{13}}.
\label{ortho}    
\eeq
Since this point is very important, let us review how the range of summation is specified, 
following e.g. ref.\cite{GS2}. 
Each generalised 6j symbol is defined so that the corresponding four pairs of screening numbers 
are non negative 
 integers. Their entries, however only differ by the spin 
with index $13$, so that 
the screening numbers  are 
$$
\Je_{1}+\Je_{3}-\Je_{13}=p_{1,3}+\pish \phat_{1,3},\quad  
\Je_{1}+\Je_{2}-\Je_{12}=p_{1,2}+\pish \phat_{1,2};
$$
$$
\Je_{13}+\Je_{2}-\Je_{123}=p_{13,2}+\pish \phat_{13,2},\quad  
\Je_{12}+\Je_{3}-\Je_{123}=p_{12,3}+\pish \phat_{12,3};
$$
\beq
\Je_{1}+\Je_{3}-{\overline \Je}_{13}=\pb_{1,3}+\pish \pbhat_{1,3},\quad 
{\overline\Je}_{13}+\Je_{2}-\Je_{123}=\pb_{13,2}+\pish \pbhat_{13,2}.
\eeq
The screening numbers are not all independent since one has 
$$
p_{1,3}+p_{13,2}=p_{1,2}+p_{12,3}=\pb_{1,3}+\pb_{13,2}, 
$$ 
with similar relations for the hatted counterparts. 
It is convenient to consider $p_{1,3}$, $\pb_{1,3}$,  $p_{1,2}$, $p_{13,2}$, and their 
hatted counterparts  as 
independent. Then clearly, only $p_{1,2}$, $\pb_{1,2}$ vary in the summation. Since 
$\Je_1$, and $\Je_2$ are fixed, we may replace the summation over $\Je_{12}$, by 
the summation over $p_{1,2}$, $\pb_{1,2}$, which is thus a discrete sum, even for 
continuous spins.     

Now, returning to Eq.\ref{phibraid1}, 
 we may resum over intermediate states on the right hand side, 
using Eq.\ref{ortho}---so that only contributions with 
$\Je_{13}=\overline \Je_{13}$ remain---and re-obtain a 
four point functions of Liouville exponentials 
 with $2$ and $3$ exchanged. One gets 
\beq 
G^{(4,\, \Phi)}_{\varpi_1, \varpi_2,\, \varpi_3,\, \varpi_4}(0, 0,\, z_2,\zb_2 ,\,
z_3,\, \zb_3,\, \infty ,\infty)=
G^{(4,\, \Phi)}_{\varpi_1, \varpi_3,\, \varpi_2,\, \varpi_4}(0, 0,\, z_3,\zb_3 ,\,
z_2,\, \zb_2,\, \infty ,\infty), 
\eeq
in agreement with locality.  Performing a similar calculation for the fusion, one gets 
a very simple result 
$$
G^{(4,\, \Phi)}_{\varpi_1, \varpi_2,\, \varpi_3,\, \varpi_4}(0, 0,\, z_2,\zb_2 ,\,
z_3,\, \zb_3,\, \infty ,\infty)=\sum_{\Je_{23}, \{\nu_{23}\}}
$$
\beq
G^{(3,\, \Phi)}_{\varpi_1, \varpi_{23},\{\nu_{23}\},\, \varpi_4,}(0, 0,\, z_1,\zb_1 ,\,
 \infty ,\infty)
G^{(3,\, \Phi)}_{\varpi_2, \varpi_{3},\, -\varpi_{23} \{\nu_{23}\},}(0, 0,\, z_3-z_1,
\zb_3-\zb_1 ,\,
 \infty ,\infty)
\eeq
Thus the fusing and braiding matrices of the Liouville exponentials are equal to one.
\section{Comments}
The summary of the operator algebra presented here shows that we have a 
perfectly consistent theory, if we restrict ourselves to screening numbers 
which are non negative 
 integers, so that condition \ref{screen} is fulfilled by 
each vertex operator. The definition of coupling constant displayed by Eq.\ref{coupl} 
is well suited for this restricted scheme, which is consistent, since the orthogonality 
relation Eq.\ref{ortho} is obtained by a discrete sum over 
integer screening charges. Assuming a symmetry between spins $J$ and
$-J-1$,   
this is sufficient to recover the results of matrix
models\cite{G5}. 
On the other hand, this symmetry leads to screening charges which are
negative integers, and in general there are arguments to include 
screening charges which are negative integers or continuous.   
In this latter case recently discussed in ref.\cite{ZZ}, by 
change of normalisation, one obtains the case discussed here as  residues of poles, 
which one may regard as on-shell contributions. Besides 
refs.\cite{DO}, and \cite{ZZ}, there are at present interesting progress 
in understanding the role of screening charges which  are non 
positive integers: see refs.\cite{S}. 
Another interesting paper\cite{T} determines the coupling constant 
directly from the conformal bootstrap of the Liouville exponential.
It seems closely related to the earlier  discussions  of
refs.\cite{CGR1}--\cite{GR1} which do so from the OPE of the chiral
components of the Liouville exponentials. 
From the viewpoint presented here, the fundamental step to include
continuous screening charges will be to generalize the orthogonality 
condition Eq.\ref{ortho}. If 6j symbols  may be 
extended to the case where conditions Eqs.\ref{1PI} are violated, so that 
a generalization of Eq.\ref{ortho} holds, then the machinary summarized
here will be at work. For continuous screening numbers, 
 Eq.\ref{ortho} would include a continuous
summation instead of a discrete one. Note that the generalized 6j symbols
with discrete screening numbers already correspond\cite{GS2}\cite{GR1}  
to the most general 
Ashkey-Wilson orthogonal polynomials presently known, so that this extension is 
highly  non trivial.    

\noindent{\bf \large \bf Acknowledgements}

\noindent 
It is a pleasure to  recall stimulating discussions with E. Cremmer,
J. Schnittger, and J. Teschner. 
 
\end{document}